\documentclass[aps,showpacs,preprintnumbers,showkeys,nofootinbib,floatfix]{revtex4}

\usepackage{amssymb}
\usepackage{amsmath}
\usepackage{graphicx}
\usepackage{graphics}
\usepackage{amsmath}
\usepackage{graphicx}
\usepackage{amssymb}
\usepackage{amsfonts}
\usepackage{latexsym}
\def\beq{\begin{eqnarray}}
\def\eeq{\end{eqnarray}}

\def\al{\alpha}
\def\be{\beta}

\def\ga{\gamma}

\def\Om{\Omega}

\begin{document}

\title{Interacting photon-baryon fluid, warm dark matter and the first acoustic peak}

\author{J\'{u}lio.C. Fabris\footnote{E-mail: fabris@pq.cnpq.br}}
\affiliation{Universidade Federal do Esp\'{\i}rito Santo,
Departamento
de F\'{\i}sica\\
Av. Fernando Ferrari, 514, Campus de Goiabeiras, CEP 29075-910,
Vit\'oria, Esp\'{\i}rito Santo, Brazil}
\author{Ilya L. Shapiro\footnote{E-mail: shapiro@fisica.ufjf.br}}
\affiliation{Departamento de F\'{\i}sica -- ICE,
Universidade Federal de Juiz de Fora,
Juiz de Fora, CEP: 36036-330, MG,  Brazil}
\author{Alan M. Velasquez-Toribio \footnote{E-mail: alan.toribio@ufes.br}}
\affiliation{Universidade Federal do Esp\'{\i}rito Santo,
Departamento
de F\'{\i}sica\\
Av. Fernando Ferrari, 514, Campus de Goiabeiras, CEP 29075-910,
Vit\'oria, Esp\'{\i}rito Santo, Brazil}
\author{Winfried Zimdahl\footnote{E-mail: winfried.zimdahl@pq.cnpq.br}}
\affiliation{Universidade Federal do Esp\'{\i}rito Santo,
Departamento
de F\'{\i}sica\\
Av. Fernando Ferrari, 514, Campus de Goiabeiras, CEP 29075-910,
Vit\'oria, Esp\'{\i}rito Santo, Brazil}

\begin{abstract}
\noindent
The Reduced Relativistic Gas (RRG) model was introduced by
A. Sakharov in 1965 for deriving the cosmic microwave background
(CMB) spectrum. It was recently reinvented by some of us to achieve
an interpolation between the radiation and dust epochs in the evolution
of the Universe. This model circumvents the complicated structure of the Boltzmann-Einstein system of equations and admits a transparent description of warm-dark-matter effects.
It is extended here to include, on a phenomenological basis, an out-of-equilibrium interaction
between radiation and baryons which is supposed to account for relevant aspects
of pre-recombination physics in a simplified manner.
Furthermore, we use the tight-coupling approximation
to explore the influence of both this interaction and of the
RRG warmness parameter
on the anisotropy spectrum of the CMB.
The predictions of the model are very similar to those
of the $\Lambda$CDM model if both the interaction and the dark-matter warmness parameters
are of the order of $10^{-4}$ or smaller. As far as the warmness parameter is concerned, this is
in good agreement with previous estimations on the basis of results from structure formation.
\end{abstract}



\maketitle

\section{Introduction}

The transition from a radiation dominated phase to a matter
dominated phase is crucial in order to understand the formation
of structures in the Universe \cite{dodelson,Mukh}. In the
standard cold dark matter (CDM) scenario, the dark matter (DM)
component decouples from the primordial plasma very early,
beginning to collapse deep in the radiative phase. This allows
to form the gravitational potential wells into which the baryons
fall after decoupling. The scenario is very successful in
predicting the large scale structure of the Universe. However,
there are some disturbing tensions at smaller scales, one of
them being the predicted large number of small structures which
do not fit observations \cite{moore,klypin}. Such tensions leave the door open for
alternative DM scenarios.
\par

One of the possibilities is to consider a warm dark-matter
(WDM) model, attributing a low, non-vanishing temperature to
the dark component \cite{DM7}. This small temperature does not
spoil the advantages of the CDM scenario at large scales but
it may, at the same time, reduce the excess of power in the spectrum
on small scales. This problem, to be treated exactly, implies
the consideration of the collisional Boltzmann-Einstein system,
including the baryon-photon interaction and the thermodynamics
of the WDM.

\par
In the present work we will develop a greatly
simplified approach that takes into account out-of-equilibrium
features of the system. To do so, we will
use the Reduced Relativistic Gas (RRG) model \cite{FlaFlu,sWIMPs}.
This model is based on the assumption that
all particles have equal kinetic energies. The use of the RRG model
substantially simplifies  the formalism, such that all
the complexity of the Boltzmann-Einstein system can be
reduced to an effective equation of state (EoS) that interpolates
between a pure radiative fluid and a pressureless matter fluid.
Remarkably, the EoS of such a system is given by a simple
algebraic formula \cite{Sakharov} (see also \cite{FlaFlu}
for a detailed derivation), which enables one to
solve the Friedmann equation exactly (for the equilibrium case) and
to obtain an explicit and transparent picture
of the
transition between the radiation phase in the early Universe and
the dust phase in the late Universe.
Indeed, the deviation from the Maxwell relativistic EoS is
very small and therefore the quality of the RRG based approximation can be evaluated as excellent \cite{FlaFlu}.

\par
Although the interaction (Thomson scattering) between baryons and photons establishes an equilibrium, equivalent to a perfect fluid description of
the combined photon-baryon system on the macroscopic level, the interaction
ceases to be effective as the decoupling era is approaching. This implies
the existence of an out-of-equilibrium period when the mean free
collision time is no longer negligible compared with the Hubble time.
We shall characterize such a period by a phenomenological out-of-equilibrium
parameter and investigate its influence on the cosmological dynamics.
Within the RRG framework we take into account temperature effects both for
the DM and for the baryons which results in (small) nonvanishing pressure
contributions of these components and we study the influence of the
corresponding ``warmness" parameters on the evolution of the Universe.
In a first step we shall find an analytic solution for the homogeneous
and isotropic background dynamics of the four-component model of
(``thermal") baryons, photons, WDM and a cosmological constant. The mere
existence of such a solution can be seen as a merit of our method
since it maps the complicated astrophysical processes of the complete
Boltzmann-Einstein system of equations on a much simpler structure. Of
course, it remains to be shown that this simplified structure really
reproduces essential features of the underlying microphysics.

\par
In a next step, using the tight coupling approximation \cite{durrer}, we look for the implications of the out-of equilibrium and warmness parameters on the position of the first acoustic peak of the CMB spectrum.
We demonstrate that the
so-called monopole mode, which defines this position, is modified due to the interaction.
Comparison with the $\Lambda$CDM model, assuming the latter grosso modo to represent a reliable reference, we obtain upper limits for the mentioned phenomenological parameters which turn out to be of the order of $10^{-4}$.
Interestingly, for the DM warmness parameter this is in agreement with previous estimations based on results for large scale structures in the universe \cite{sWIMPs}.

\par
The paper is organized as follows. In the next section, Sec.~\ref{basic}, we
construct and work out the equations for the coupled system of
baryons and radiation. The balance equations for our four-component model are solved exactly which
provides us with an explicit expression for the Hubble parameter in terms of the scale factor.
In Sec.~\ref{tight} we use the tight-coupling approximation to
study the influence of the interaction and warmness parameters on the position of the first acoustic peak of the CMB spectrum.
Finally, in the last section, Sec.~\ref{conclusion} we draw our conclusions and
discuss further perspectives of the RRG model.

\section{Basic equations of the interacting RRG model}
\label{basic}

We consider a four-component cosmic model consisting of baryons,
photons, DM and a cosmological constant. Both baryons and DM are
described as a relativistic gas of massive particles. Furthermore,
we include an interaction between baryons and photons in a
phenomenological manner.
Microscopically, photons and baryons interact via Thomson scattering which establishes
an equilibrium between them. As a consequence, both components are treated as perfect fluids
with the same temperature. Here, we take into account, in a phenomenological manner, the possibility of small deviations from this equilibrium.
Moreover, the baryon pressure, although small, is not assumed to be zero exactly.

The dynamics of the photon-baryon
system is then described by the following system of equations:
\beq
\frac{d\rho_{b}}{dt} + 3H(\rho_{b}+P_{b})  &=& \gamma_{rb} \rho_{r}
- \gamma_{br}\rho_{b}\,,
\\
\label{1}
\frac{d\rho_{r}}{dt} + 3H(\rho_{r}+P_{r})&=& - \gamma_{rb} \rho_{r}
+ \gamma_{br}\rho_{b}\,.
\label{2}
\eeq
Here, $\rho_{b}$ and $\rho_{r}$ are the energy densities of
baryonic matter and of radiation, respectively, while
$P_{b}$ and $P_{r}$ are the corresponding pressures.
The quantities $\gamma_{rb}$ and $\gamma_{br}$ denote the rates
by which energy is transferred from radiation to baryons and from
baryons to radiation, respectively, and $H = a^{-1}da/dt$ is the Hubble rate with $a$ being the scale factor of the Robertson-Walker metric.
In the state of equilibrium  one has
\beq
\label{3}
\overline{\gamma}_{rb} \rho_{r} - \overline{\gamma}_{br}\rho_{b}
= -\left(\overline{\gamma}_{br}
- \frac{\rho_{r}}{\rho_{b}}\,\overline{\gamma}_{rb}\right)\rho_{b}
= 0 \,,
\eeq
where the bars indicate the equilibrium values of the corresponding quantities. Deviations from equilibrium can be mapped onto a single
constant parameter $\xi$ according to the simple approximation
\beq
\label{4}
\gamma_{br} - \frac{\rho_r}{\rho_b}\,\gamma_{rb} = \xi H \,.
\eeq
In this case, the relevant set of basic equations can be written as
\beq
&& \frac{d\rho_b}{dt} + 3H(\rho_b + P_b)
\,=\, -\xi H \rho_{b}\,,
\label{5}
\\
&& \frac{d\rho_r}{dt} + 3H(\rho_r + P_r)
\,=\, \xi H \rho_{b}\,.
\label{6}
\\
&& \frac{d\rho_{D}}{dt} + 3H(\rho_{D}+P_{D})
\,=\, 0\,.
\label{7}
\\
&& \frac{d\rho_{\Lambda}}{dt} \,=\, 0\,,
\label{8}
\eeq
where $\rho_{D}$ and $P_{D}$ are energy density and pressure, respectively, of the
DM component and \ $\rho_{\Lambda}$ \ is the density of the dark
energy (cosmological constant, in our case).

The pressures of the warm components and radiation
in the above equations are given by \cite{FlaFlu}
\beq
P_{i}
&=& \frac{\rho_{i}}{3}\,\Big(1-\frac{\rho^2_{di}}{\rho_{i}^2}\Big)\,,
\label{9}
\\
P_{r} &=& \frac{\rho_{r}}{3}\,,
\label{10}
\eeq
 where $i=b,D$, i.e., $i$
corresponds to baryonic matter or dark matter, respectively, and
$\rho_{di}=\rho_{i1}(1+z)^{3}$
is the mass (static energy) density.
Let us stress that the main RRG relation (\ref{9}) reproduces
the EoS of the relativistic Maxwell distribution with a very
good precision and can be used as a reliable and
simple approximation for describing the warm matter components
in the Universe \cite{Sakharov,FlaFlu}. The new aspect of
the present work is the interaction which we introduced
phenomenologically in Eqs. (\ref{5}) and (\ref{6}).

It is easy to see that the Eq. (\ref{5}) can be solved independently
of Eq.  (\ref{6}). Using (\ref{9}), we can cast the equation (\ref{5})
in the form of a Bernoulli differential equation which can be easily
solved to give
\beq
\rho_{b}(a)
&=& \sqrt{\frac{\rho_{b1}^2}{1+\xi}\,a^{-6}
+ \Big(\rho_{b0}^{2}-\frac{\rho_{b1}^2}{1+\xi}\Big)a^{-2(4+\xi)}}
\label{11}
\\
\nonumber
\\
 &=& \frac{\rho_{b1}}{\sqrt{1+\xi}} \,a^{-3} \,
 \Big[1 + b^2 a^{-2 -2 \xi}\Big]^{1/2}\ ,
\label{12}
\eeq
where $$
 \quad
\rho_{b1} = \frac{\rho_{b0}\sqrt{1+\xi}}{\sqrt{1+b^{2}}}
\quad \mbox{and} \quad
b^{2}=\frac{\rho_{b0}^{2}}{\rho_{b1}^{2}}(1+\xi)-1\,.
$$
Here, $\rho_{b0}$ and $\rho_{b1}$ are integration constants which
have a clear physical interpretation \cite{FlaFlu} in case of
$\xi=0$. For the present moment, with $a=1$, we have
\ $\rho_{b}(1) = \rho_{b0}$, while the ratio between
 $\rho_{b1}$  and  $\rho_{b0}$  measures the warmness
of the baryonic matter constituent.
 The same role is played by the parameter  $b$
in a different parametrization. An interaction term $\xi \neq 0$ just renormalizes
the corresponding values.

For $\xi=0$ we consistently recover the ideal
relativistic gas RRG case from (\ref{12}). In this limit the solution
is a square root of the sum of the squares of the dust-like
and radiation-like terms.
Notice that this form is different from the simple
sum of the dust and radiation components. In order to see this
explicitly, consider the case when the dust component is
dominating, that means $\rho_{b0}(1+z)^3 \gg \rho_{r0}(1+z)^4$.
Then we can rewrite eq. (\ref{12}) as
\beq
\rho_{ideal}(a) &=& \frac{\rho_{b0}}{\sqrt{1+b^{2}}} a^{-3}
\Big[1 + b^2 a^{-2}\Big]^{1/2} \approx
\frac{\rho_{b0}}{\sqrt{1+b^{2}}} a^{-3}
+ \frac{\rho_{b0}b^{2}}{2\sqrt{1+b^{2}}}\, a^{-5}\,.
\label{13}
\eeq
Obviously, the last
term in (\ref{13}) has a scaling behavior which is distinct
from the one of the radiation with a small dust component. It
is easy to see that at the intermediate stages the difference
is even greater. Indeed, Eq.~(\ref{11}) shows that for $\xi =0$
the gas is close to radiation for a very large positive red-shift
and to dust when the red-shift approaches $-1$. One can see that
the relativistic gas is cooling down with the expansion of the
Universe, such that its radiation-like part  becomes weaker.

For $\xi > 0$ we have a similar
behavior in the distant past but, as to be expected, the
relativistic gas cools down faster and the radiation component is
decreasing less rapidly than in the ideal gas case. Physically
this means the gas of massive particles heats up the radiation.

On the contrary, for $\xi < 0$ the equation (\ref{11})
indicates to an opposite effect. The relativistic gas of
massive particles is absorbing energy from the radiation
and cools down slower compared to the ideal gas case.
Moreover, starting from some negative value of $\xi$
the gas may not cool down at all and even start to heat
up when the Universe expands.

Using the solution (\ref{12}) in equation (\ref{6})
for the  radiation component, the latter takes the form
\beq
&&\frac{d\rho_{r}(a)}{da}+\frac{4\rho_{r}}{a}
= \frac{\sqrt{1 + b^2\,a^{-2-2\xi}}}{\sqrt{1 + b^{2}}}
\,\times\,\frac{\xi \,\rho_{bm0}}{a^4}\,,
\label{14}
\eeq
which has an analytic solution
\beq
f_{r}(a) \,=\, \frac{\rho_{r}(a)}{\rho_{r0}} &=& \Big[1+\frac{\rho_{b0}}{\rho_{r0}}\,G(\xi,b,1)\Big]a^{-4}
- \frac{\rho_{b0}}{\rho_{r0}}\,\, G(\xi,b,a)\,,
\label{15}
\eeq
where $\rho_{r0}$ is the present value of $\rho_{r}$ and the
function $\,G(\xi,b,a)\,$ is defined as
\beq
G(\xi,b,a) &=& \frac{a^{-3} }{\sqrt{1+b^{2}}}
\left[(1+\xi)\,_2F_{1}(\al,\be,\ga,\zeta)
\,-\, \sqrt{1+b^{2}a^{-2-2\xi}}\right]\,.
\label{16}
\eeq
Here, $\,_2F_{1}(\al,\be,\ga,\zeta)\,$ is the hypergeometric function,
which has a branch-cut discontinuity in the complex $\zeta$ plane
running from \ $1$ \ to \ $\infty$ \ and is defined as
\beq
_{2}F_{1}(\alpha,\beta,\gamma,\zeta) \,=\, \sum_{k=0}^{\infty}\frac{(\al)_{k}(\be)_{k}}{(\ga)_{k}}\,
\frac{\zeta^{k}}{k!}\,,
\label{17}
\eeq
where $(\al)_{k}$ is the Pochhammer symbol. For our case we find
that
\beq
&&\alpha=-\frac{1}{2 + 2 \xi},\\
\label{18}
&&\beta= \frac{1}{2}\ ,\\
\label{19}
&&\gamma=1-\frac{1}{2 + 2\xi},\\
\label{20}
&&\zeta = - \frac{b^{2}}{a^{(2 + 2\xi)}}.
\label{21}
\eeq

When $\xi=0$ then $-\alpha=\beta-\ga=\frac{1}{2}$ and in this case there
is a simple form for the hypergeometric function, i.e. $(1-\zeta)^{\frac{1}{2}}$
\cite{hf}, where $\zeta$  is given by  equation (\ref{21}). Thus the solution
for $\xi =0$ corresponds to equation (2)
of reference \cite{fsv}. Finally, the equation (\ref{7}) for the DM energy
density is decoupled from the other components, and its solution is given by
\beq
f_{D}(a)=\frac{\rho_{D}}{\rho_{D0}} = \frac{a^{-3}}{\sqrt{1+b_{1}^{2}}}\sqrt{1+\frac{b_{1}^{2}}{a^{2}}}\
\label{22}
\eeq
with $b_{1}^{2}=(\rho_{D0}^{2}-\rho_{D1}^{2})/\rho_{D1}^{2}$.

Combining equations (\ref{12}), (\ref{15}) and (\ref{22}) and restricting ourselves to the spatially flat case,
the Hubble parameter for our model is explicitly given  by
\begin{eqnarray}
E(a) &=& \frac{H^{2}}{H_{0}^{2}}
\,=\, \big[
\Omega_{b0}f_{b}(a) + \Omega_{r0}f_{r}(a) + \Omega_{D0}f_{D}(a)
+ (1-\Omega_{b0}-\Omega_{r0}-\Omega_{D0})\big]\,,
\label{23}
\end{eqnarray}
where the $\Omega_{i0}$ (now here $i = b, r, D, \Lambda$)
represent the ratios of the present-time values of the energy densities and the
critical energy density.

\begin{figure}[htb]
\begin{center}
\includegraphics[height= 6.0 cm,width=9.0cm]{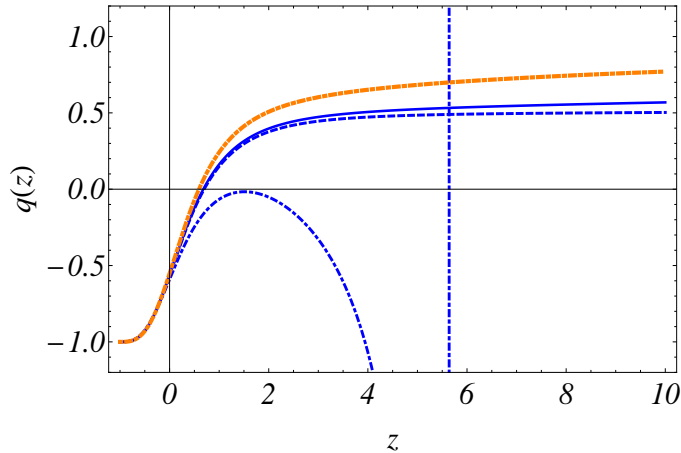}
\end{center}
\caption{Redshift dependence of the deceleration parameter $q(z)$ for
different values of $\xi$. From top to bottom: $\xi=0.7,0.1,0,-0.7$.
In general, for redshifts $z \gtrsim 2$ the value of the deceleration
parameter tends to $q \approx 0.5$, which is characteristic of an era
dominated by matter. The transition redshift for a wide range of values
of the $\xi$ parameter lies in the region $z <1$. As mentioned in the
text, for negative values of $\xi$  the
gas is absorbing energy from the radiation. For high negative values the
model exhibits a singular behavior as, e.g., in the case shown here. In
all cases we have used the values
$\Omega_{b0}=0.04$, $b=0.001$, $b_{1}=0.01$  and $\Omega_{D0}= 0.25$. Notice
that for a better qualitative visualization we have chosen much higher values
of $|\xi|$ than admitted by our analysis in Sec.~\ref{tight}.}
\label{fig1}
\end{figure}

\begin{figure}[htb]
\begin{center}
\includegraphics[height= 5.0 cm,width=11.0cm]{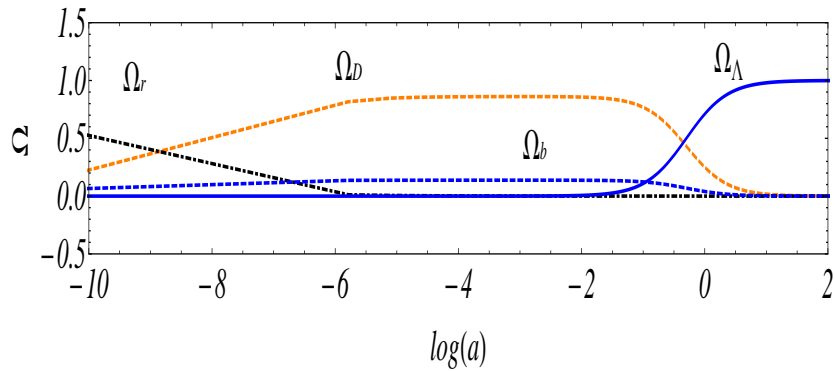}
\end{center}
\caption{Dependence of the density parameters on the scale factor for all four components.
This figure shows the transition from a radiation-dominated phase to a DM-dominated phase as well as a subsequent
transition to a final period where the cosmological constant dominates.
In all case we have used the values
$\Omega_{b0}=0.04$, $b=0.001$, $b_{1}=0.01$, $\xi=10^{-2}$
and $\Omega_{D0}= 0.25$.}
\label{fig1-1}
\end{figure}

It is useful to characterizes the dynamics of
our model with the help of the redshift dependence of the deceleration parameter
\beq
q(z) &=& \frac{1+z}{H}\, \frac{dH}{dz}\,-\,1\,, \qquad z = \frac{1}{a} - 1\,,
\label{24}
\eeq
which results in the plots of Fig.~\ref{fig1}.\\
On the other hand, the fractional density parameters for
arbitrary times are defined as
\beq
\Om_{i} = \frac{\rho_{i}}{\rho_{c}}
= \frac{8 \pi G \rho_{i}}{3H^{2}}\,.
\label{25}
\eeq
These density parameters are plotted in Fig. \ref{fig1-1} for all four components.

Let us emphasize that our model admits analytical solutions for
the entire homogeneous and isotropic background dynamics, including interaction and warmness effects. This should  definitely be a
very welcome feature for the sake of reconstruction of the history
of the Universe by using observational data. Along with practical
advantages of analytic expressions, it is well-known that, in
the use of numerical solutions, any additional derivative or
integration results in new correlations and this increases the
error in the final result. This aspect is important in both
parametric and nonparametric approaches. For details on this
issue see references \cite{hols}.

\section{The Tight Coupling Approximation}
\label{tight}

In general, the study of anisotropies
 in the CMB requires to use a
system of thousands of coupled equations \cite{dodelson} (Boltzmann
hierarchy). However, progress has been made by implementing numerical
codes as CAMB  \cite{lewis}. Even so, to study the implications of a
given cosmological model for the CMB is a task that involves
the Boltzmann equations together with the perturbed Einstein equations.
This objective is beyond the scope
of our paper. Instead, we shall resort to the tight-coupling approximation which
we believe to be a reasonable simplification in the present context.

Thus, we assume that before recombination, photons and baryons
are tightly coupled since Thomson scattering happens much faster
than the expansion of the Universe. Quantitatively, this is described in terms of the
optical depth $\tau$,
\beq
\tau &\equiv& \int_{\eta}^{\eta_{0}}{d\eta' n_{e} \sigma_{T} a}
\,\,\gg \,\,1\,,
\label{26}
\eeq
where $n_{e}$ is the electron number density and $\sigma_{T}$
denotes the Thomson cross section.
Originally this approximation was implemented by Peebles and Yu
\cite{py} (see also Hu and Sugiyama \cite{sh}).

Following Ref. \cite{dodelson},
the only
nonnegligible momenta $\Theta_{l}$ in the Boltzmann hierarchy in the limit $\tau \gg 1$
are the monopole $(l=0)$
and the dipole $(l=1)$. All the higher momenta are suppressed.
As a result one obtains an equation
for the density contrast with the help of which it is possible
to derive an expression for the position of the first acoustic peak.
In the standard model this position is well determined by the
fit given by Hu and Sugiyama \cite{sh}.
Although it is strictly valid only for the $\Lambda$CDM model, we can use this fit her as well because
values $\xi<10^{-3}$ are suggested from observational constraints
on the sound horizon $r_{s}$. This is shown in Fig. 3, where
we have included measurements of $r_{s}$ made by WMAP \cite{wmap} and
Planck \cite{planck}. For values $\xi> 0.05$
the sound horizon is outside the observational limits for a wide range
of values of the  matter density parameters.
In this context it is important to note that in our approach the influence of the
interaction parameter $\xi$ on the first acoustic peak is entirely due to the dependence of
the Hubble parameter on $\xi$.

In order to make the presentation clear, let us recall the derivation
of the equation for the density contrast in the tight-coupling approximation.
We shall follow here reference \cite{md} and use the uniform curvature gauge. 
Then, the perturbation equations for the baryons are given by \cite{md}
\beq
&& \dot{D_{b}} \,=\, -kV_{b}\,,
\label{27}
\\
&& \dot{V_{b}} + H V_{b}
\,=\, k\Psi + \frac{\dot{\tau}}{R}(V_{r}-V_{b})\,
\label{28}
\eeq
where $V_k$ and $D_k$ are gauge invariant velocity and density 
perturbations for the fluid $k$ (we use notations of \cite{durrer1}), 
over-dot indicates a derivative with respect
at the conformal time $\eta$ and $\dot{\tau}=an_{e} \sigma_{T}$ is
the differential optical depth. Quite similarly, the equations for 
the photons are
\beq
&& \dot{D_{r}} \,=\, -\frac{4}{3} k V_{r}\,,
\\
\label{29}
&& \dot{V_{r}} \,=\,
2k\Psi + \frac{1}{4} k D_{r} - \dot{\tau}(V_{r}-V_{b})\,.
\label{30}
\eeq
The relations between the variables $D$ and  $V$ of the uniform curvature gauge
and those of the longitudinal gauge are given by
\cite{durrer1}
\beq
D_{b,r}&=&\delta^{long}_{b,r} - 3(1+w_{b,r})\Psi\,,
\\
\label{31}
V_{b,r}&=&v^{long}_{b,r}\,,
\label{32}
\eeq
where the superscript ``long" refers to the longitudinal gauge, $\delta^{long}_{b,r}$ are the corresponding fractional density perturbations for baryons and photons, respectively, and the quantities $w_{b,r}$ denote their EoS parameters.
The velocity potentials $v^{long}_{b,r}$ are related to the
four-velocities by
\beq
U^{\alpha}_{b,r} &=& \big((1-\Psi),\,v^{long ,i}_{b,r}\big)\,
\nonumber
\eeq
 and $\Psi$ is the
 Newtonian potential. The last terms on the right-hand sides of (\ref{28})
 and (\ref{30}) can be associated with the collision term for Thomson scattering of the
 Boltzmann equation. The
 details of this derivation can be followed in the reference
 \cite{mb}.
Now we rewrite (\ref{30}) as
\beq
V_{r}-V_{b} &=&
\frac{2k}{\dot{\tau}}\,
\Big(\Psi + \frac{D_{r}}{8}\Big)
\,-\, \frac{1}{\dot{\tau}}\,\dot{V_{b}}\,.
\label{35}
\eeq
The tight-coupling regime is characterized by a high rate of
collision between baryons and photons. Therefore, an
expansion with respect to $\dot{\tau}^{-1}$ is a reasonable approximation.
In zeroth order, we get
\beq
V_{b} = V_{r}\, \quad  \Rightarrow \quad \dot{V}_{b} = \dot{V}_{r}\,.
\label{33}
\eeq
This is the first step of an iteration approach, first presented by Peebles and Yu \cite{py}.
Furthermore, using Eq.~(\ref{33})
in Eqs. (\ref{27}) - (\ref{29}) we get
\beq
D_{r} &=& \frac{4}{3} D_{b}\,.
\label{36}
\eeq
This relation characterizes an adiabatic evolution.
Since $\xi \ll 1$, the adiabatic approximation is justified.
In fact, due to the small value of \ $\xi$, all non-adiabatic
contributions, typical for an interacting model, become
negligible at the first-order approximation. Of course, for
larger  \ $\xi$ \ the situation can be different.

By using the equations (\ref{33}-\ref{36}) in equation
(\ref{28}), we arrive at
\beq
\ddot{D_{b}} + \frac{R}{1+R} H \dot{D_{b}} + \frac{k^{2}}{3(1+R)}D_{b}
\,=\, -\frac{2+R}{1+R} k^{2} \Psi\,.
\label{38}
\eeq
\noindent
To achieve the common form given in the literature for the above
equation, the speed sound should be defined as
\beq
c_{s}^{2} &=& \frac{1}{\sqrt{3(1+R)}}\,,
\label{39}
\eeq
where $R$ is the photon-baryon momentum-density ratio
that can be written as \cite{mb}
\begin{equation}
R=\frac{(P_{b}+\rho_{b})V_{b}}{(P_{r}+\rho_{r})V_{r}} = \frac{P_{b}+\rho_{b}}{P_{r}+\rho_{r}}
\approx \frac{3\rho_{b}}{4\rho_{r}}
\label{37}
\end{equation}

Due to the presence of the pressure $P_{b}$, the ratio $R$ in (\ref{37}) does not exactly coincide with the standard ratio $R = (3/4)\rho_{b}/\rho_{r}$. Numerically, is possible show that the difference between both expressions is less than $10\%$ and we shall use the approximation in the last part of (\ref{37}) in the following.\\
\ \\
{Equation (\ref{38}) with (\ref{39}) and (\ref{37}) is the second-order differential equation for a forced, damped harmonic oscillator which governs the acoustic
oscillations of the photon-baryon fluid.
The oscillation period is determined by the
sound speed and hence by the baryon and photon densities. In our
case it is given by the solutions (\ref{11}) and (\ref{15})
for baryons and photons respectively. Via these solutions, the interaction parameter $\xi$
influences the sound speed.

To solve Eq.~(\ref{38}),
we suppose $R$ to be slowly varying over an oscillation period inside
of the sound horizon.
Making use of the WKB method \cite{dodelson},
we obtain the general solution, which can be written as
\beq
D_{b}(k,\eta) &=&
D_{b0} \,\Big(\frac{1}{1+R}\Big)^{1/4} \cos(kr_{s})
\,-\, E(k,\eta)\,,
\label{42}
\eeq
where
\beq
E(k,\eta) &=& (1+R)^{-1/4} \int_{0}^{\eta}d\be
\,\Big[
\frac{2+R}{(1+R)^{3/4}} \,\,
\frac{\sin \left[k r_{s}(\eta)-kr_{s}(\beta)\right]}{kc_{s}}\,\,
k^{2} \Psi\Big]\,.
\label{43}
\eeq
In the limit when the first term in Eq.~(\ref{42}) dominates,
the peaks and troughs should appear at the extremals of
\ $\cos(kr_{s})$. \ Following the references \cite{hue,dodelson,durrer},
the location of the first peak is conveniently fit as
\beq
k_{1,peak} \,=\, \frac{5\pi}{2r_{s}}\, \big(1+0.217 \Omega_{D}h^{2}\big)\,.
\label{44}
\eeq
The sound horizon $r_{s}$ at decoupling, which appears in Eqs.~(\ref{42}) and (\ref{43}), is defined as the
comoving distance that a wave can travel prior to decoupling:
\beq
r_{s} &=& \int_{0}^{a_{dec}}{\frac{c_{s}\,da}{a^2\,H(a)}}
\,=\, \int_0^{a_{dec}}{\frac{da}{a^{2} H(a) \sqrt{3(1+R)}}}\,.
\label{45}
\eeq
Here,  $a_{dec}$ is the scale factor at the
time of decoupling.

In Fig. 3. we depict the sound horizon at
decoupling as a function of the interaction parameter $\xi$
for three different values of the matter density parameter.
One can see that values less than $\xi \leq 0.5 \times 10^{-2}$
are numerically compatible with observational constraints of the
last dataset of PLANCK \cite{wmap}: \ $r_{s}=144\pm 0.71$.
Furthermore, Fig. 4. shows that the difference between
\ $\Lambda CDM$ \ and our model is very small for a value of
$\xi = 10^{-3}$. However, the difference increases ​for a greater
value of the interaction parameter.

\begin{figure}[htb]
\begin{center}
\includegraphics[height= 10.0 cm,width=15.0cm]{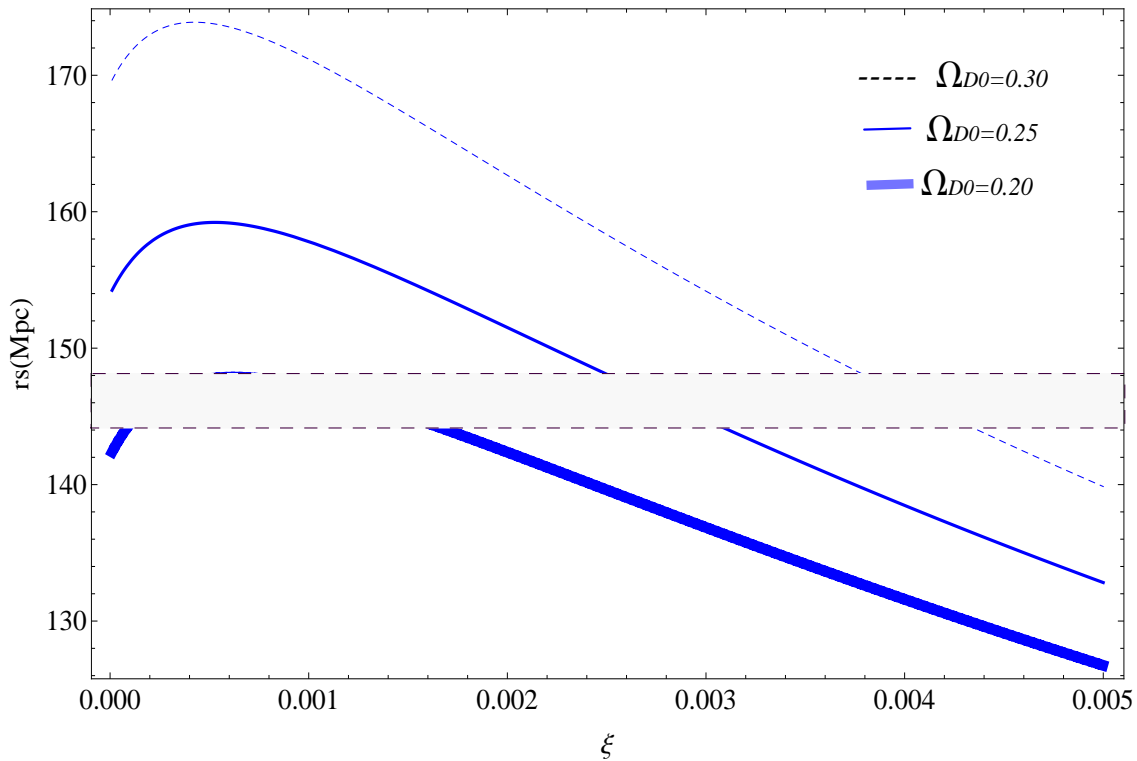}
\end{center}
\caption{The sound horizon at decoupling as function of the
interaction parameter. The two horizontal lines show the observational constraints
given by the measurements of PLANCK and WMAP \cite{wmap,planck}. We have
multiplied the error by two to be conservative. In all figures we used
$b_{1}=0.0001$ and $\Omega_{b0}=0.05$.\\
}
\label{fig1-2}
\end{figure}

\begin{figure}[htb]
\begin{center}
\includegraphics[height= 7.0 cm,width=7.0cm]{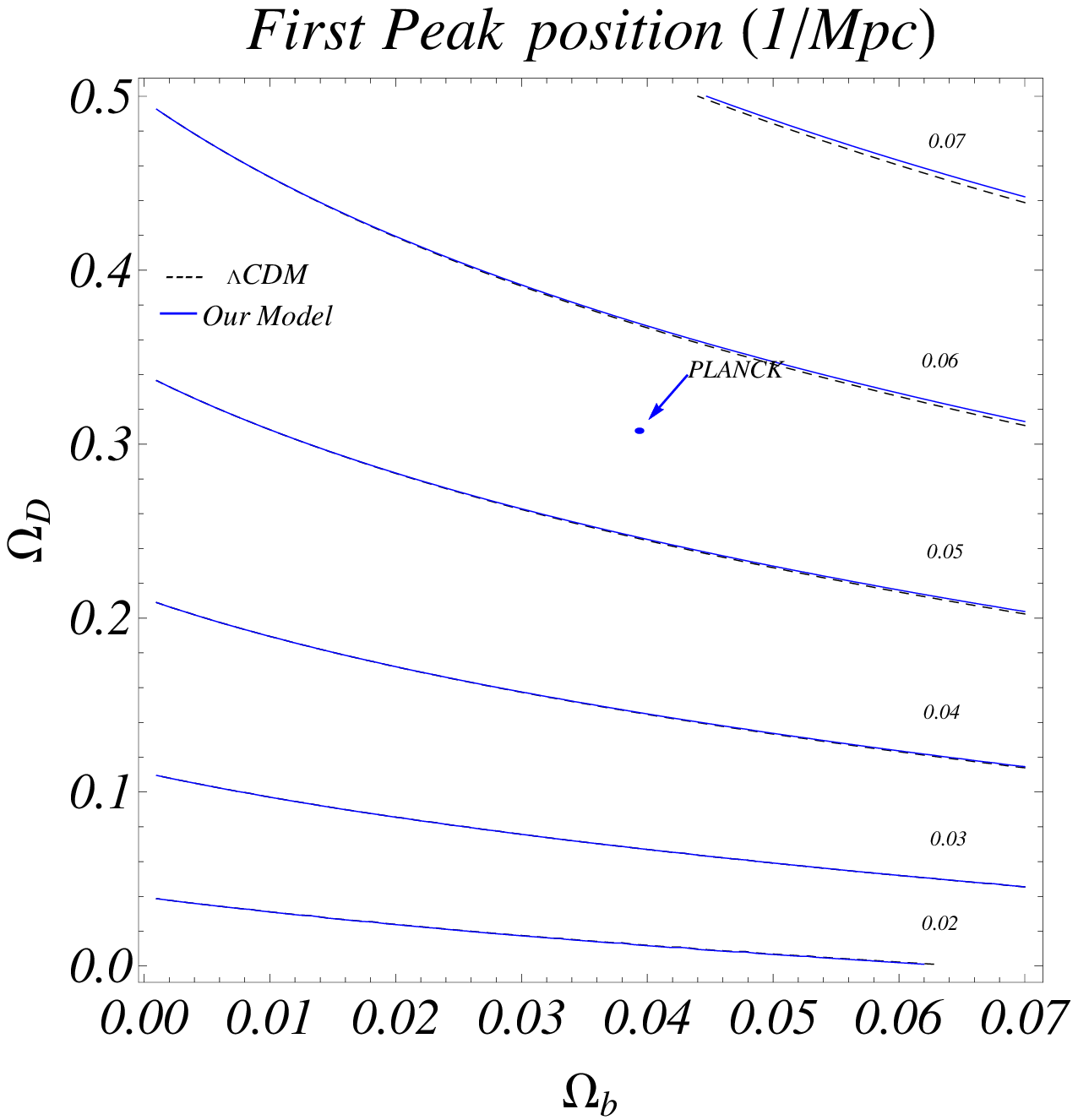}
\includegraphics[height= 7.0 cm,width=7.0cm]{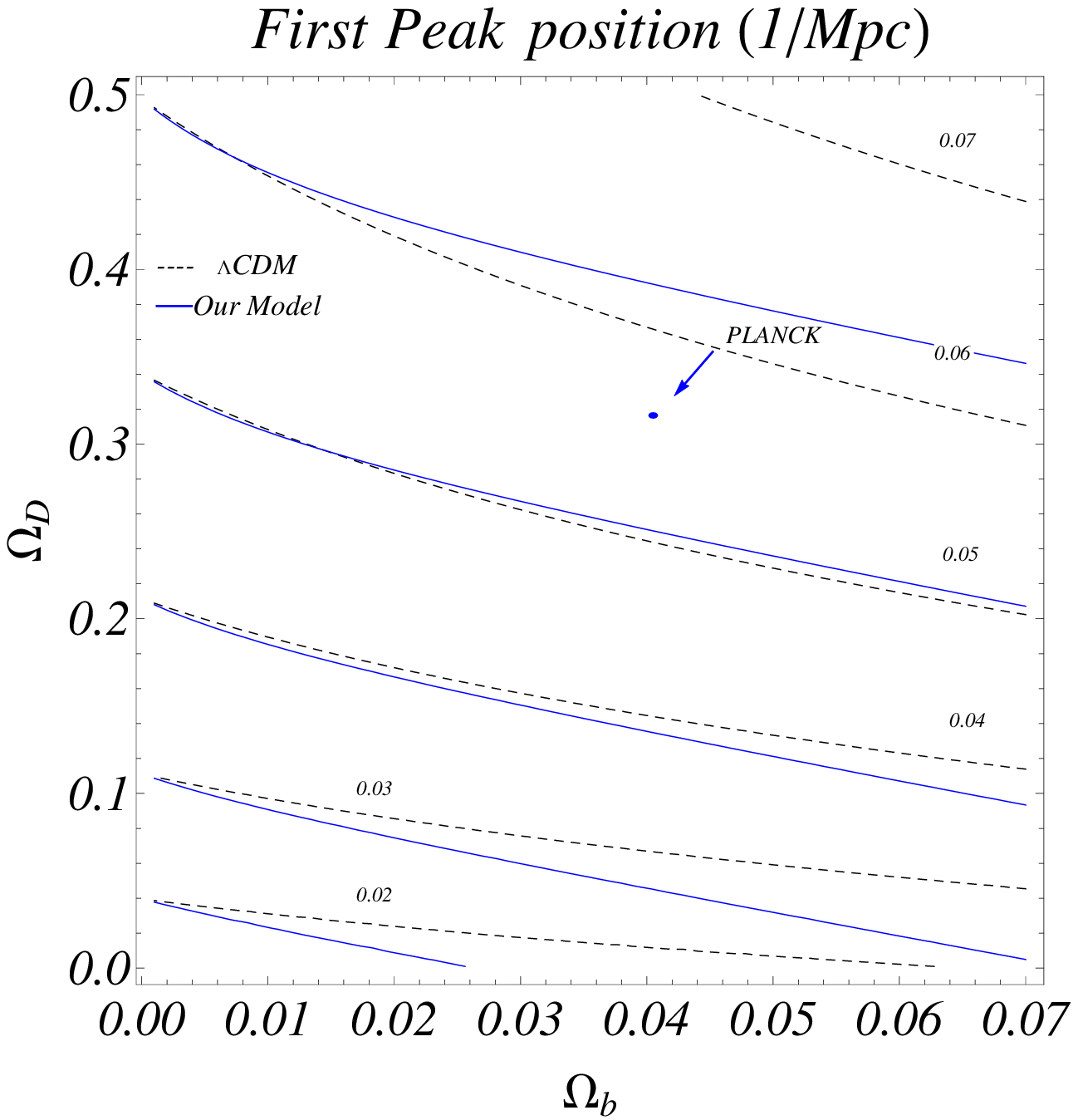}
\includegraphics[height= 7.0 cm,width=7.0cm]{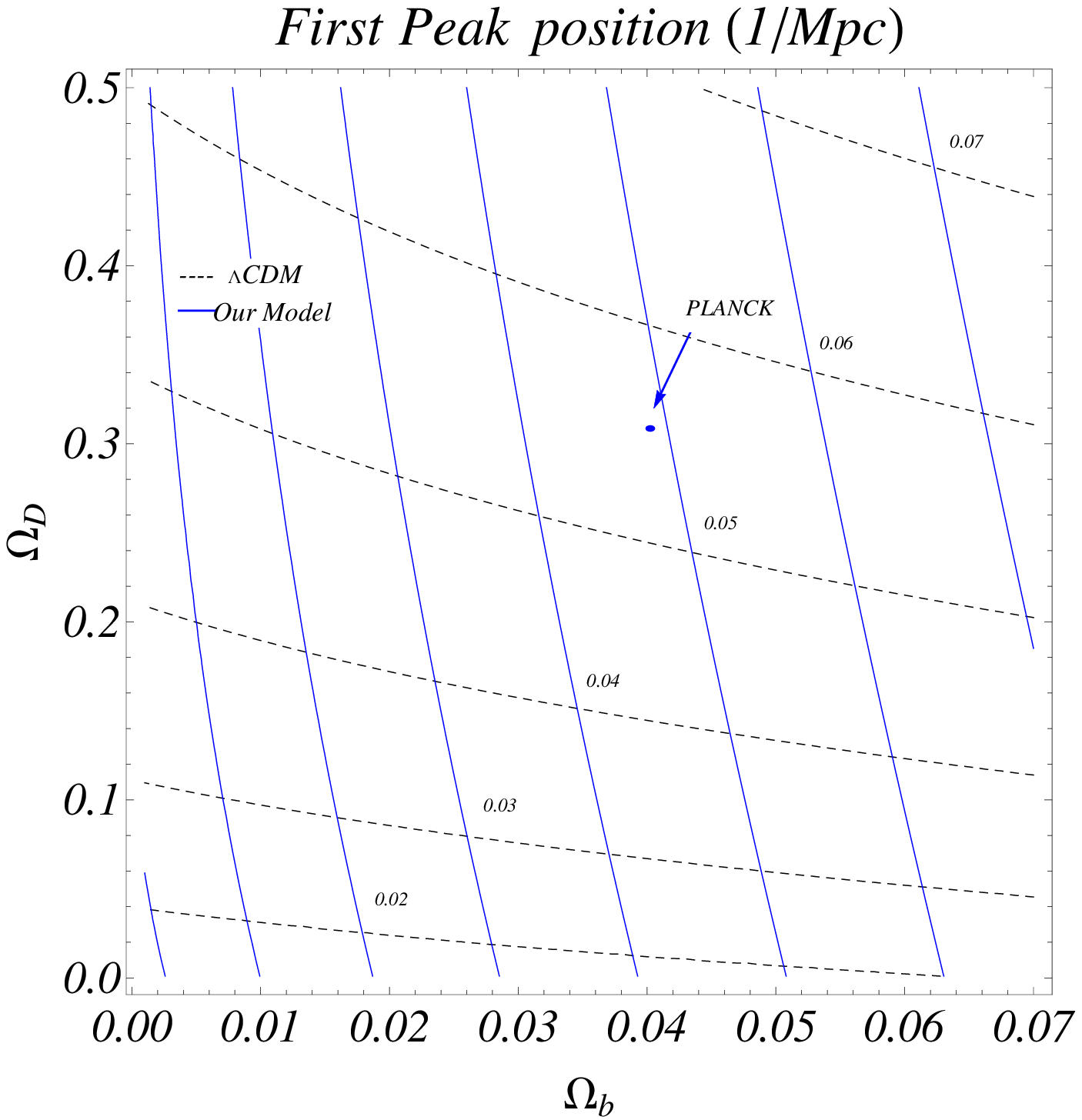}
\includegraphics[height= 7.0 cm,width=7.0cm]{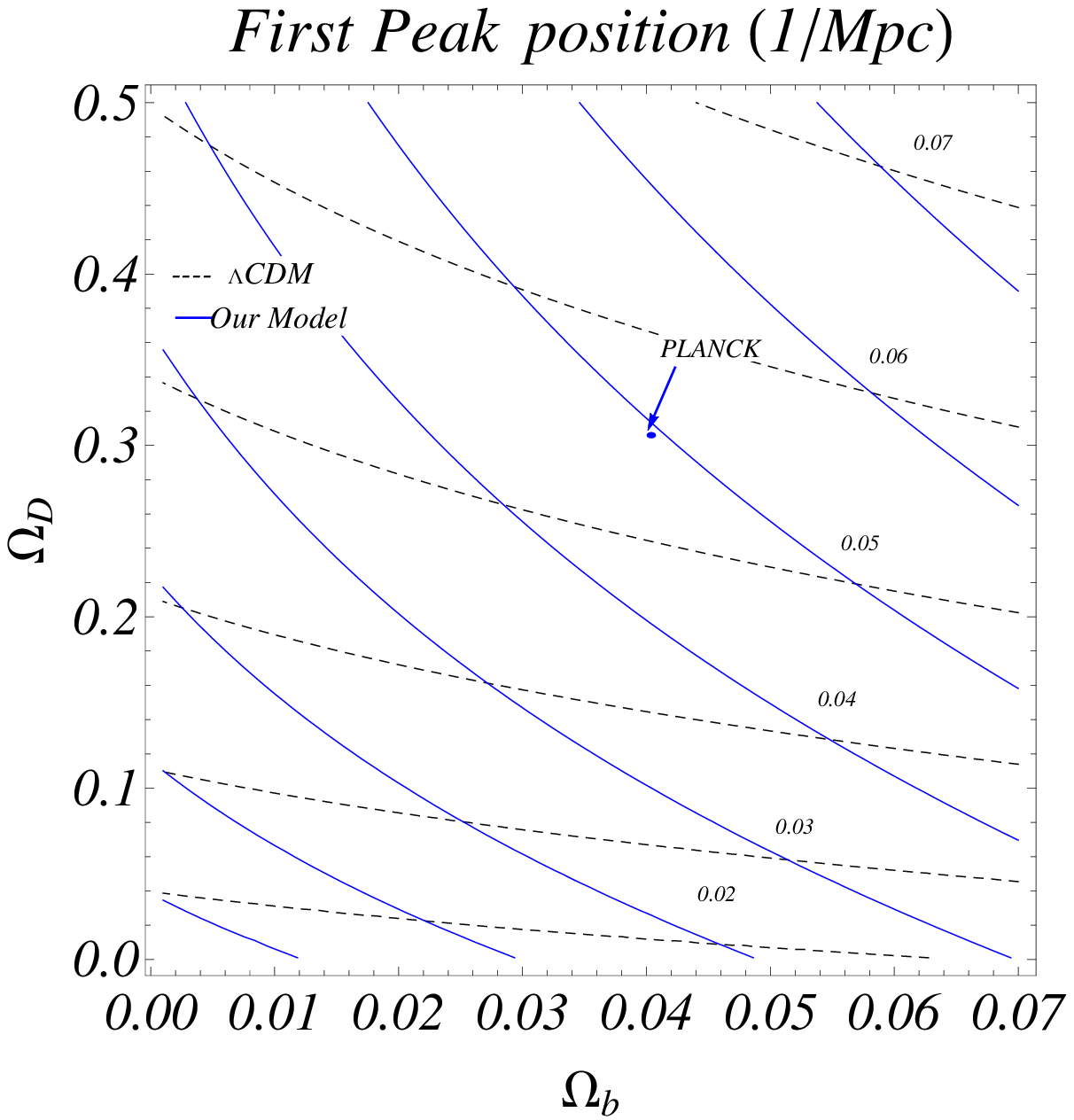}
\end{center}
\caption{The position $k_{1,peak}$ of the first acoustic peak (cf. Eq.~(\ref{44})) as function of the density parameters.
The solid lines represent our model and the dashed lines
the $\Lambda$CDM model. For the values $\xi \approx 10^{-4}$, $b \approx 10^{-4}$ and
$b_{1} \approx 10^{-4}$ (upper left panel) our model is indistinguishable from the
$\Lambda$CDM model. When the value of $\xi$ increases to $10^{-3}$ (upper right panel) the difference between the models is already evident.
In the lower panels we consider the influence of the WDM parameter
$b_{1}$ for the fixed values $\xi=10^{-4}$ and
$b=10^{-4}$. For $b_{1}=10^{-2}$ for example
(bottom left), the departure from the $\Lambda$CDM model is dramatic, implying a high degeneracy of the
$\Omega_{D0}$ parameter.
Already for $b_{1}= 0.5 \times 10^{-3}$(bottom right) the differences are substantial.
In all case we used $h=0.7$ and the point represent the
measure from PLANCK ($\approx 0.055$) \cite{planck, wmap} with best fit of the density parameters given by
$\Omega_{\Lambda}=0.685^{+0.018}_{-0.016}$ and $\Omega_{D0}=0.315^{+0.016}_{-0.018}$.}
\label{fig1-3}
\end{figure}

\section{Conclusions}
\label{conclusion}

We generalized a previously constructed, simplified RRG-based cosmological model \cite{Sakharov,FlaFlu}, made of WDM, a cosmological constant, baryons and photons by taking into account an out-of-equilibrium interaction within the baryon-photon fluid. Such interaction, characterized here by a single phenomenological parameter $\xi$, is supposed to be relevant before decoupling when the scattering rate ceases to be much higher than the Hubble rate and deviations from equilibrium are expected.
We found an exact analytic solution for the homogeneous and isotropic background which  encodes the impact of this parameter on the dynamics as well as warmness effects of both DM and baryons. This solution interpolates the cosmic evolution from an early radiation-dominated phase, followed by a transition to matter dominance until a final de Sitter stage.

In a second step, using the tight-coupling approximation, we considered perturbations in the photon-baryon fluid on this background and studied the influence of the out-of equilibrium and warmness parameters on the position of the first acoustic peak of the CMB spectrum. We found that both the parameter $\xi$ and the DM warmness parameter $b_{1}$ have to be of the order of $10^{-4}$ or less to be compatible with observational data and with the $\Lambda$CDM model. As far as $\xi$ is concerned, this can be seen as a confirmation of the perfect-fluid approach for the interacting photon-baryon system since deviations from equilibrium do not seem to be important.
If $b_{1}$ considerably exceeds the value $10^{-4}$ there is a degeneracy in the DM density such that almost all values of the DM parameter $\Omega_{D0}$ respecting the flat condition are compatible with $\Omega_{b0} \sim 0.05$. The restriction on $b_{1}$ is in agreement with the results obtained for the equilibrium RRG model using the large scale structures data \cite{sWIMPs}. On the other hand, even a small degree of warmness may potentially be useful to cure problems of the CDM paradigm, such as the cusps in the density profiles of galaxies and the excess of galactic satellites \cite{moore,klypin}. An important procedure to break the degeneracy with the $\Lambda$CDM model for $b_{1} \leq 10^{-4}$ is to inspect the non-linear regime. This implies to adapt the usual computations used for CDM to the case where there is a departure from coldness, equivalent to the appearance of a pressure component. We hope to perform a corresponding analysis in future work.\\



\end{document}